\documentclass[preprint]{aastex}
\shorttitle{[0~III] Emission in BH Host GC RZ2109}
\shortauthors{Steele et al.}

\newcommand{\gtsim}{\ {\raise-0.5ex\hbox{$\buildrel>\over\sim$}}\ }
\newcommand{\ltsim}{\ {\raise-0.5ex\hbox{$\buildrel<\over\sim$}}\ }
\newcommand{\kms}{km\thinspace s$^{-1}$}

\def\Msun{\hbox{$\thinspace M_{\odot}$}}

\begin{document}
\title{Velocity Structure and Variability of [O~III] Emission in Black Hole Host
Globular Cluster RZ2109}

\author{Matthew M. Steele\altaffilmark{1}, Stephen
  E. Zepf\altaffilmark{1},Arunav Kundu\altaffilmark{1,2}, Thomas
  J. Maccarone\altaffilmark{3}, Katherine L. Rhode\altaffilmark{4}, and John
  J. Salzer\altaffilmark{4}}

\altaffiltext{1}{Department of Physics \& Astronomy, Michigan State University, East Lansing, MI 48824; e-mail: steele24@msu.edu}

\altaffiltext{2}{Eureka Scientific, 2452 Delmer Street Suite 100, Oakland, CA 94602-3017}
\altaffiltext{3}{School of Physics \& Astronomy, University of Southampton,
  Southampton, Hampshire S017 1BJ, UK}

\altaffiltext{4}{Department of Astronomy, Indiana University, Bloomington, IN 47405}

\begin{abstract}
We present a multi-facility study of the optical spectrum of the extragalactic
globular cluster RZ2109, which hosts a bright black hole X-ray source.  The
optical spectrum of RZ2109 shows strong and very broad
[O~III]$\lambda\lambda$4959,5007 emission in addition to the stellar absorption
lines typical of a globular cluster.  We use observations over an extended
period of time to constrain the variability of these [O~III] emission lines. We
find that the equivalent width of the lines is similar in all of the datasets;
the change in L[O~III]$\lambda$5007 is $\ltsim 10\%$ between the first and last
observations, which were separated by 467 days.  The velocity profile of the
line also shows no significant variability over this interval. Using a simple
geometric model we demonstrate that the observed [O~III]$\lambda$5007 line
velocity structure can be described by a two component model with most of the
flux contributed by a bipolar conical outflow of about 1,600 \kms , and the
remainder from a Gaussian component with a FWHM of several hundred \kms.

\end{abstract}


\section{INTRODUCTION}

The first unambiguous black hole X-ray source in a globular cluster was
discovered by Maccarone et al.\ (2007) in the extragalactic globular cluster
RZ2109. This globular cluster is in the Virgo elliptical galaxy NGC~4472 (M49),
and is a luminous, metal-poor cluster, with an absolute magnitude of M$_V =
-10.0$ and a $B$-$V$ color of $0.68$, located $6.6'$ away from the center of
its host galaxy \citep{RZ01}.  The black hole X-ray source in RZ2109 was
discovered in XMM observations of the NGC~4472 globular cluster system, of
which RZ2109 is a member.  The XMM observations of the source
XMMU122939.7+075333 in RZ2109 showed that it had a X-ray luminosity of $\simeq
4 \times 10^{39}$ ergs~s$^{-1}$ with x-ray counts that varied by a factor of 7
over a few hours \citep{Maccarone07}. The high X-ray luminosity, more than an
order of magnitude higher than the Eddington luminosity for a neutron star,
requires either a black hole or multiple neutron stars in the old population of
the globular cluster. The strong variability within the XMM observation rules
out multiple neutron stars, and indicates that the source is a black-hole
system \citep{Maccarone07}.

A study of existing optical spectra of RZ2109 revealed that it had broad,
strong [O~III]$\lambda$5007 emission at the absorption line radial velocity of
the globular cluster \citep{Zepf07}.  Follow-up optical spectroscopy with
the Keck telescope showed stellar absorption lines typical of an old globular
cluster and remarkably broad and luminous [O~III]$\lambda \lambda$4959,5007
emission lines \citep[hereafter Z08]{Zepf08}. Specifically, Z08 showed
that the velocity width of the [O~III]$\lambda$5007 line is $\simeq 2,000$
km~s$^{-1}$, and that the line luminosity is about $1.4 \times 10^{37}$
ergs~s$^{-1}$. Moreover, there is no sign of emission lines other than
[O~III], and the L([O~III])/L(H$\beta$) ratio appears to be at least 30 (Z08).

The observed high luminosity and broad velocity width of the [O~III]$\lambda
\lambda$4959,5007 emission lines place strong constraints on the origin of
these emission lines and the nature of the black hole X-ray source in RZ2109.
For example, the presence of this very strong and broad [O~III] emission in the
black hole X-source hosting globular cluster RZ2109 and the absence of similar
emission in any other spectroscopic study of globular clusters argues strongly
against significant beaming of the X-rays \citep{Gnedin2009}.  Furthermore, Z08
showed that the observed broad [O~III]$\lambda\lambda$4959,5007 velocities
cannot be explained by gravitational motions near the black hole because the
available volume sufficiently close to the black hole is many orders of
magnitude too small to produce the observed [O~III]$\lambda$5007, given the
critical density of the line. Instead, the broad velocity and high luminosity
requires a strong outflow from the accreting black hole.  Z08 suggests that
such strong outflows occur for systems near their Eddington limit \citep[see
  also][and references therein]{Proga2007}.  The observed L$_{X}$ of $\simeq 4
\times 10^{39}$ ergs~s$^{-1}$ thus indicates a stellar mass for the accreting
black hole in the globular cluster RZ2109 (see discussion in Z08).

In this study we present several sets of new optical spectroscopic data on the
broad [O~III]$\lambda\lambda$4959,5007 lines in RZ2109.  We use the multiple
observations at different times to study the variability of the [O~III]$\lambda
\lambda$4959,5007 emission over baselines ranging from one to 467 days. We
also take advantage of the higher signal-to-noise and higher spectral
resolution of our new Gemini data to constrain models of the [O~III]$\lambda
\lambda$4959,5007 emitting regions.

\section{OBSERVATIONS AND DATA REDUCTION} \label{sec:obs}
In this study we use medium resolution optical spectra of RZ2109 
from four observatories. One set of observations has been previously 
presented, and the other three are described for the first time here.
The four sets of observations span an interval of 467 days from the 
initial to the most recent observation.

\subsection{Keck}
Optical spectra were obtained using the Low Resolution Imaging Spectrograph
\citep{Oke95} on the Keck Telescope.  These data, originally reported in
Z08, have a wavelength coverage of 3200--5500, 5800--8930 \AA\ and a
measured spectral resolution of R $\sim400$.  The spectra were collected on the
nights of UT 2007 December 17--18.
 
\subsection {WHT}
Observations were made on UT 2008 January 5--6 using the 4.2 m William Herschel
Telescope (WHT) with the Intermediate dispersion Spectrograph and Imaging
System operating in longslit mode. The WHT spectra have a wavelength coverage
of 3900--5350, 5390--10000 \AA\ and a measured resolution of R $\sim 660$. To
produce these observations the spectrograph was set up using the 1.53 arcsec
slit with the R300B grating in the blue arm and the R158R grating in the red
arm.  For purposes of calibration the standard star HZ44 was observed using an
identical instrumental setup.

\subsection{SOAR}
Observations from Goodman High Throughput Spectrograph \citep{Clemens04} on the
4.1m Southern Astrophysical Research Telescope (SOAR) were collected on UT 2009
January 23, 24 and February 22. The SOAR spectra have a wavelength coverage of
4352--7024 \AA\ and a measured resolution of R $\sim 1600$, using the KOSI 600
grating and a slit width of 0.84 arcsec.  
  
\subsection{Gemini}
Spectra from the Gemini South Telescope were obtained under a queue observation
using the Gemini Multi-Object Spectrograph \citep{Hook04} in longslit mode.
The data were obtained on the nights of UT 2009 March 28, 29 and 30 (program
GS-2009A-Q-1).  The instrumental setup included the use of the B1200 grating
with a slit width of 1.0 arcsec. This setup produced a wavelength coverage for
the resulting spectra is 4445--6023 \AA\ with a measured spectral resolution of
R $\sim 2400$.

\subsection{Data Reduction}

All data presented for the first time in this work were reduced using standard
IRAF NOAO longslit tools, including the Gemini IRAF package for the Gemini
data. The processed two dimensional spectra were extracted and, except where
noted in Section \ref{sec:lpv}, the data from the same facility obtained on
consecutive nights were co-added to increase the signal-to-noise ratio of each
observation.  The low signal-to-noise of the SOAR observations due to the
smaller instrument aperture and relative inefficiency of the spectrograph at
the time necessitated co-addition of the two dimensional spectra prior to
extraction of the one dimensional spectrum. The WHT spectrum was flux
calibrated using observations of standard star HZ44; all other spectra were
left uncalibrated and simply normalized to a polynomial continuum fit. For the
[O~III] line profile measurements and modeling presented in this work, simple
continuum normalization is sufficient.  Flux calibration is useful for
analysis of the stellar component of the spectrum, where the detailed flux of
the continuum contains information on the cluster's stellar population.

The stellar component of the globular cluster is fitted with a synthetic
stellar spectrum as described in Section \ref{sec:scm}. The best fit model is
used to account for and remove any effect of the overall stellar population of
the globular cluster on the measurement of the emission line equivalent width
in Section \ref{sec:ew}.  The [O~III] emission line complex is then analyzed
for variability in Section \ref{sec:var} and the velocity profile is modeled in
Section \ref{sec:gm} to investigate the geometry of the emitting region.

\subsection{Stellar Component Model} \label{sec:scm}

The observed spectra presented in this work are a superposition of the emission
line system of interest and the emission from stars of the host globular
cluster. In order to analyze the emission line system we first fit and subtract
the stellar component.  In fitting the stellar component we follow the
prescription of \citet{Koleva2008}. A grid of synthetic spectra was created
using the Pegas\`e-HR code of \citet{LeBorgne2004} and the Elodie 3.1 spectral
library \citep{Prugniel2007}. The models assumed a Salpeter initial mass
function (IMF) and a single epoch of star formation in the cluster. The
synthetic spectra were fit using the WHT data due to its large wavelength
coverage and moderate instrumental resolution. The WHT data were masked to
exclude the region of [O~III]$\lambda \lambda 4959,5007$ and H$\beta$ to
eliminate any effects of emission lines. The data were then fitted using the
grid of spectra which yielded a best fit model with an age of 12 Gyr and an
initial [Fe/H] of -1.2.  In order to test this model fitting scheme an
independent set of synthetic stellar population models from \citet{Vazdekis}
were re-sampled and injected with noise to match our data, and then fit using
the method described above.  This test yielded estimates of the uncertainty in
the stellar component parameters contributed by the model fitting scheme of
$\pm1.5$ Gyr in age and $\pm0.05$ dex in metallicity.  Given the uncertainties
in the stellar population models and the assumptions used in the generation of
the synthetic spectra, we adopt total uncertainties of age $\pm2$ Gyr and
metallicity $\pm0.1$ dex.  Repeating the stellar component fitting while
adopting an IMF for \citet{Kroupa2001} produces fit stellar parameters of 14
Gyr and [Fe/H] of -1.1. Both these parameters are within the estimated
uncertainties of the modeling and fitting procedure, suggesting the effect of
IMF selection on the resulting stellar component model is minimal for this
study.  In \citet{Maccarone07} a value of [Fe/H]=$-1.7 \pm 0.2$ and an old
stellar population age was inferred from the optical colors of
\citet{RZ01}. These parameters are in reasonable agreement with the values we
find here.  A more detailed analysis of the stellar component of the globular
cluster will be presented in a forthcoming work (Steele et al. 2011, in prep).

The stellar population parameters determined above were used to construct
synthetic spectra to match the wavelength coverage and spectral 
resolution of each observation.  These synthetic spectra were then
used to remove the stellar component from each observation.
 
\section{VARIABILITY} \label{sec:var}

A careful study of the variability of the [O~III] complex equivalent width and
velocity structure, when combined with the x-ray variability timescales,
potentially provides information on the size and density structure of the
emitting region. By examining the equivalent width of the [O~III] over various
timescales it is possible to constrain the ionizing photon path length and
estimate the size of the emitting region.  The emission line velocity
structure, coupled with any variability in X-ray data, may also help constrain
other physical properties of the emitting material.

\subsection{Equivalent Width Variability} \label{sec:ew}

\begin{figure}
  \begin{center}
    \plotone{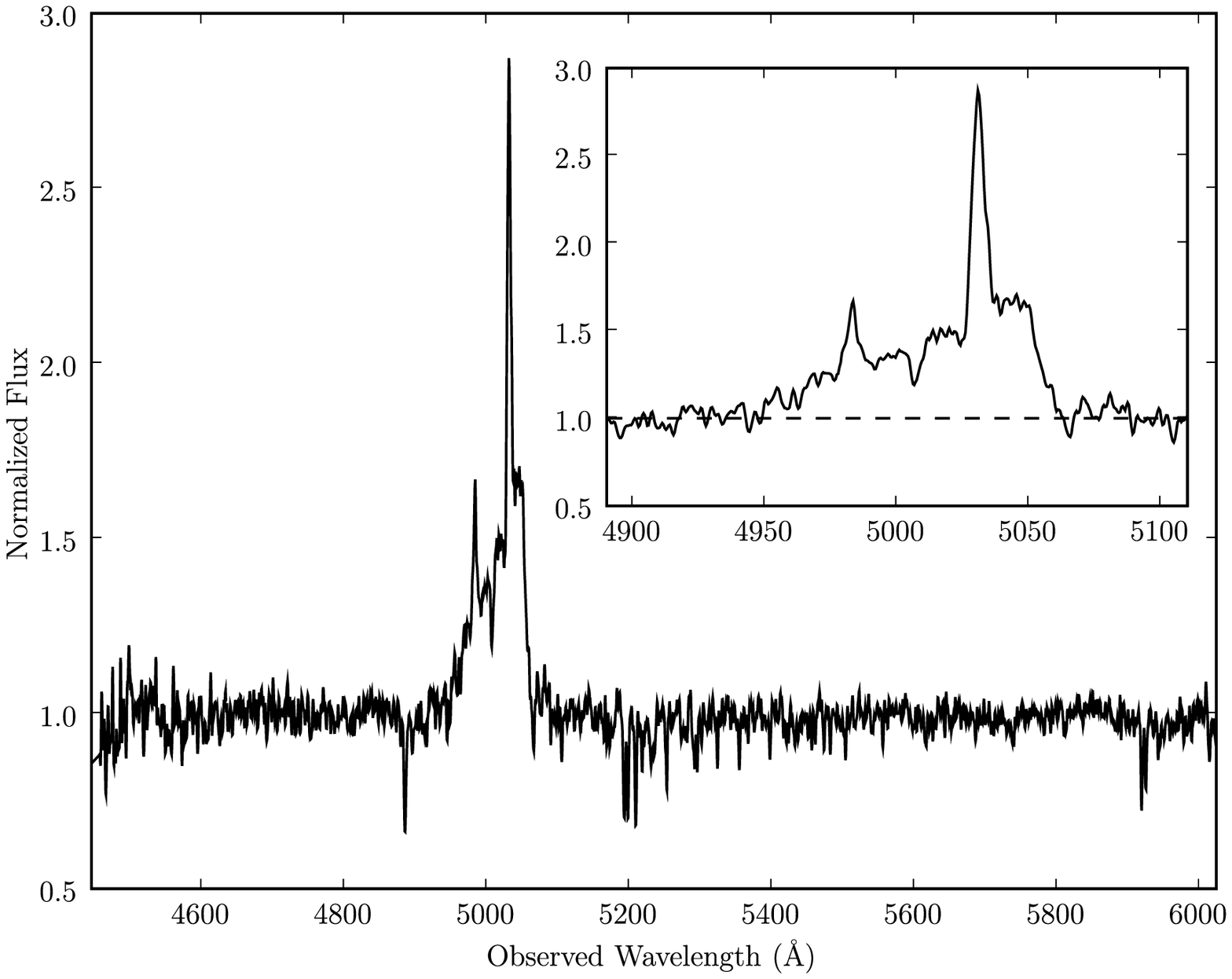}
    \caption{Gemini GMOS spectrum of black hole hosting globular cluster
      RZ2109. The inset displays the [O~III]$\lambda \lambda$4959,5007 emission
      structure which dominates the spectrum. The spectrum has been continuum
      fit and normalized, and smoothed with a three pixel boxcar function.}\label{fig:gmos}
  \end{center}
\end{figure}

Line equivalent widths were measured by direct integration under the line
profile and bounded by a linear fit to the local continuum.  The
[O~III]$\lambda \lambda$4959,5007 emission is observed as a blended feature,
due to the very broad width of each [O~III] line, as seen in Figure
\ref{fig:gmos}. The [O~III]$\lambda \lambda$4959,5007 emission lines and the
stellar absorption of the globular cluster RZ2109 are also found to have a
redshift of 1475 km~s$^{-1}$ \citep{Zepf07}.  Therefore, the equivalent width
of the entire complex was measured between the observed wavelengths of 4964
\AA\ and 5058 \AA , which are symmetric in velocity space about the midpoint of
the redshifted broad [O~III]$\lambda\lambda$4959,5007 emission and beyond which
the emission flux is indistinguishable from zero.  The equivalent widths of the
individual $\lambda$5007 and $\lambda$4959 lines were then set by the ratio
established by their respective transition probabilities. Identical
measurements were performed on the stellar component models and the resulting
values subtracted from the data measurements to yield equivalent widths of the
emission component. Uncertainty estimates for the equivalent widths were
generated by considering a systematic contribution (instrumental noise, poor
cosmic ray subtraction, etc.)\ and a photon statistic contribution.  The
systematic contribution was estimated by measuring the per pixel
root-mean-squared (RMS) deviation from the continuum of two regions located to
the red and blue side of the emission feature which were free of strong stellar
absorption features. The two uncertainty contributions were added by quadrature
to produce the given values.
 
Table \ref{tab:ew} displays the measured equivalent widths of the
[O~III]$\lambda$5007 for each observation along with the date of observation.
The uncertainty-weighted mean of all observation is $33.4 \pm 0.3$ \AA\ with
little change observed over the course of our observations.  For the longest
time baseline between the first observation with the Keck telescope to the last
with Gemini-South, we find very close agreement in the [O~III]$\lambda$5007
emission, with a a difference of only $\sim 9 \pm 2\% $ between these
observations separated by 467 days. The intervening WHT and SOAR data have
larger uncertainties, but no observation deviates strongly from the mean
[O~III] equivalent width, with the WHT observation being the most offset with a
formal difference of $27.5 \pm 7.5\% $ from the weighted mean.

\subsection{Line Profile Variability}\label{sec:lpv}

\begin{figure}
  \begin{center}
    \plotone{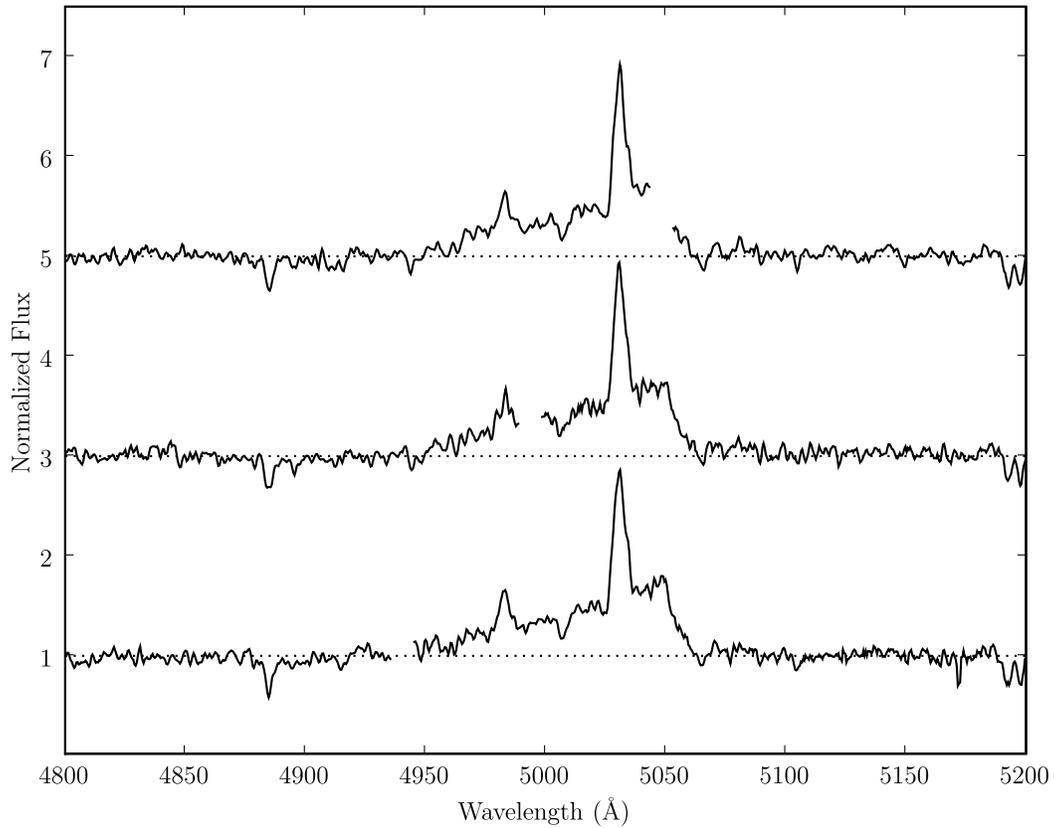}
    \caption{Gemini spectra internight variability. The lower (March 28, 2009),
      middle (March 29, 2009), and upper (March 30, 2009) spectra have been
      smoothed with a three pixel boxcar function and offset on the flux axis
      for legibility. The dashed lines give the level of the normalized
      continuum for reference.}\label{fig:inv}
  \end{center}
\end{figure}

It is possible to examine the internight variability of the [O~III] complex
velocity profile using the Gemini spectra co-added for each of the three
consecutive nights of observation. In Figure \ref{fig:inv} the three two hour
exposures are displayed. A visual inspection of the three spectra does not
reveal any obvious variability over the region displayed. To test the line
profile variability each spectrum was divided into 5 \AA\ bins and subtracted
from the other two binned spectra. The differences in the bins in the [O~III]
region were then calculated in terms of the RMS of the differences in bins not
containing [O~III] emission. Of the 57 comparison bins in the [O~III] region, 17
displayed differences larger than 1$\sigma$, two larger than 2$\sigma$ and none
larger than 3$\sigma$. For these differences, seven of the 17 1$\sigma$
deviations were above the continuum and ten were below it. These numbers are
consistent with the RMS statistical expectations, indicating no evidence for
variability in the velocity structure over the three nights observed.

\begin{figure}
  \begin{center}
    \plotone{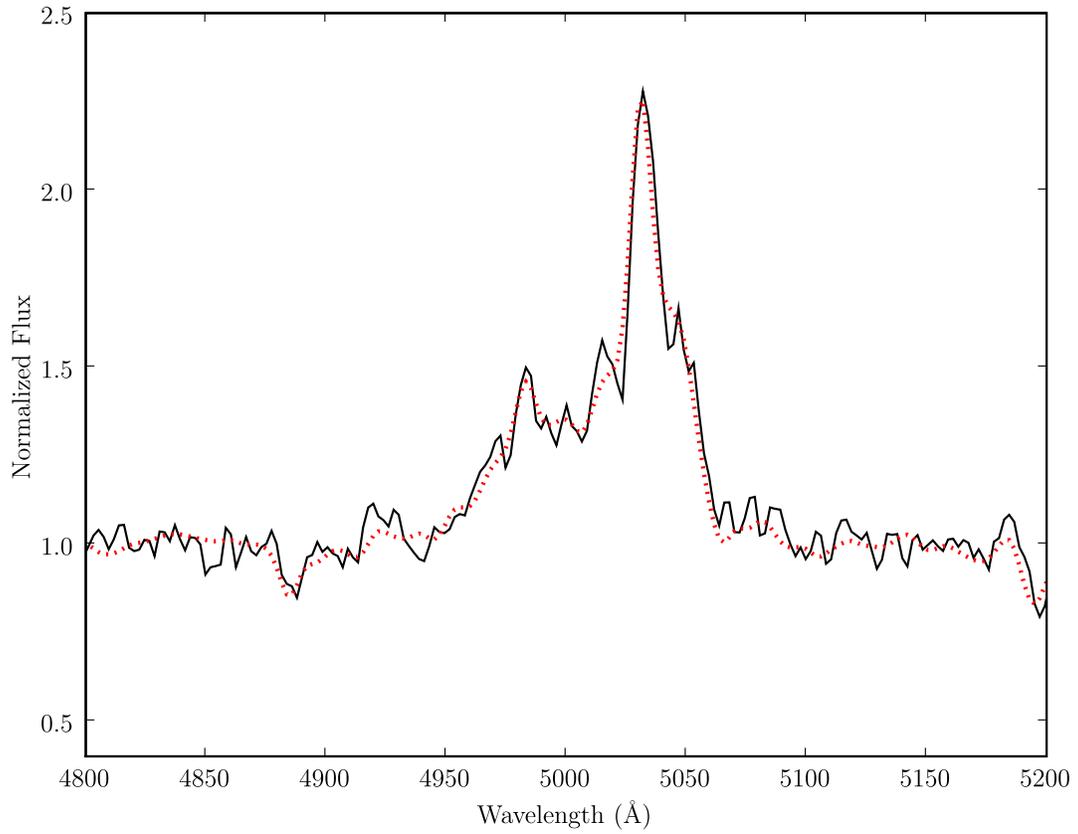}
    \caption{Keck and Gemini spectra comparison. The Gemini spectra (dotted
      line) have been convolved with the instrumental resolution of the Keck
      spectra (solid line.)  The Keck observation preceded the Gemini
      observation by 467 days.}\label{fig:kvg}
  \end{center}
\end{figure}

A similar approach may be used to probe a longer temporal scale by a
comparison of the Keck and Gemini spectra. For this analysis the six hours of
Gemini integrations are co-added and convolved with the Keck instrumental
resolution. The Keck and convolved Gemini spectra are displayed in Figure
\ref{fig:kvg}.  A visual inspection reveals that when the resolution of the two
spectra are matched, the observations are consistent.  A similar analysis as
performed on the Gemini individual night spectra yields 11 bins in the [O~III]
emitting region, four of which have differences greater than 1$\sigma$, one
greater 2$\sigma$ and none the greater than 3$\sigma$.  These results are again
consistent with the RMS statistical expectations.  The 10 \AA\ bin with the
greatest difference of 2.4 $\sigma$ is centered at 5063 \AA\ , on the extreme
red side of the emission complex.  By eye it is possible to pick out a small
excess of emission in what appears to be a 'secondary peak' on the blue side of
both the [O~III]$\lambda$4959 and [O~III]$\lambda$5007 structures in the Keck
spectrum.  However when the $\lambda$4959 and $\lambda$5007 lines are
transformed to velocity space these 'secondary peaks' are separated by 330 km
s$^{-1}$, suggesting they are not real structures.

\section{GEOMETRIC MODEL}\label{sec:gm}
\begin{figure}
  \begin{center}
    \plotone{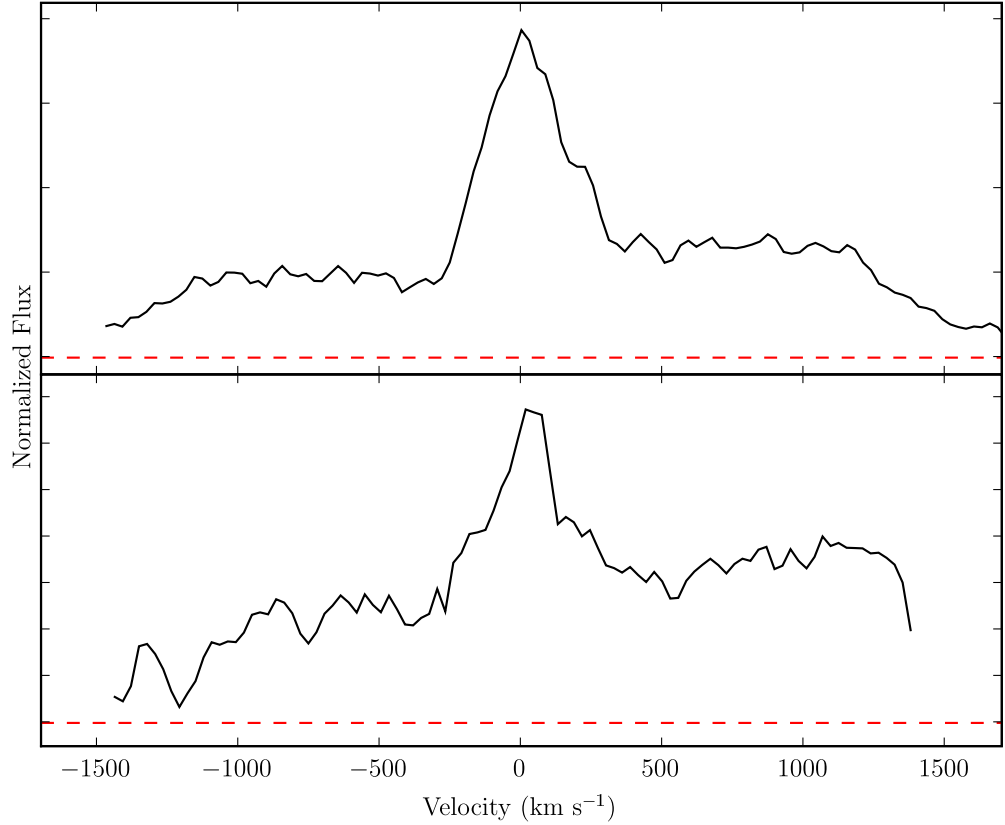}
    \caption{Gemini spectra [O~III] velocity structure. The top figure displays
      the line profile for [O~III]$\lambda 5007$, and the lower panel
      [O~III]$\lambda 4959$. The two lines have been separated by a cut at an
      observed wavelength of $\lambda 5006$. As a result there may be a small
      degree of blending on the extreme blue side of the [O~III]$\lambda 5007$
      and the extreme red side of [O~III]$\lambda 4959$.  The dashed lines give
      the continuum level for reference.  }\label{fig:prof}
  \end{center}
\end{figure}

An initial qualitative inspection of the [O~III]$\lambda \lambda 4959,5007$
velocity structure line profile reveals that the complex may be divided into a
$\lambda 4959$ line and a $\lambda 5007$ line shown in Figure \ref{fig:prof}.
Theoretical considerations dictate that these two lines should display identical
line profile shapes and differ only in their total flux level, the ratio of
which is set by their transition probabilities. The data are in good
agreement with this expectation.  The two main features of the [O~III] line profiles are strong broad
emission spanning between $\sim$300--1300 km~s$^{-1}$, and a central, narrower
peak.  At larger velocities the emission falls off rapidly between $\sim
$1300--1700 km~s$^{-1}$.  While the velocity structure appears symmetric about
line center, there appears to be a flux level asymmetry between the blue and
red sides of the line profile.

In an effort to quantify the observed emission line profile, we fit a simple
geometric model to the [O~III]$\lambda$5007 line profile of the continuum
normalized Gemini data. The model consists of two components; a bipolar conical
outflow and a lower velocity width Gaussian component.  The bipolar conical
outflow produces the broad emission component, with the lower velocity width
peak contributed by the Gaussian. This two component model is necessary as no
model with only a bipolar conical component or only a Gaussian component was
able to simultaneously reproduce flat broad component, and the narrower peak.
Fitting the simple geometric model to the observations allows us to describe
the emission line structure in terms of the model parameters: the velocity of
the outflowing gas, the opening angle of the cones, the angle of inclination of
the cone's axis relative to the observer, and the full width half maximum
(FWHM) of the Gaussian component.

\begin{figure}
  \begin{center}
    \plotone{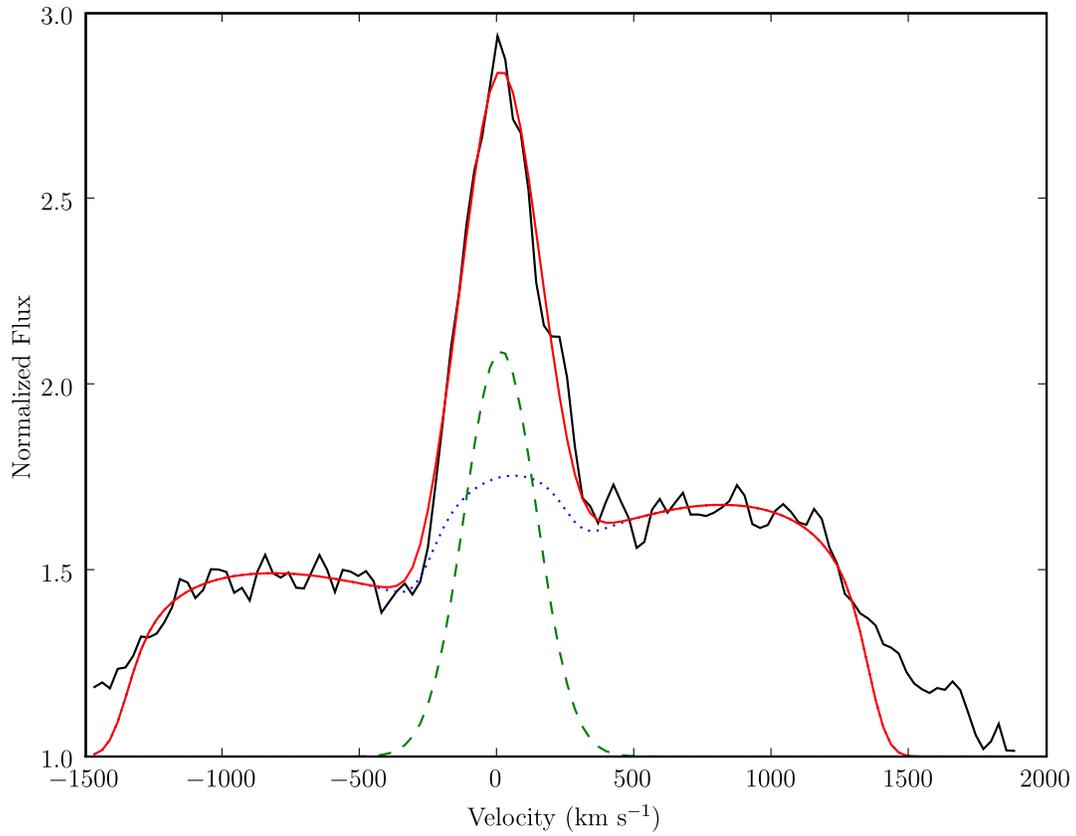}
    \caption{Gemini spectra with geometric model.  The solid black line shows
      the Gemini spectra, the dotted line displays the best fit bipolar conical
      outflow component of the geometric model, the dashed line the Gaussian
      component of the model, and the solid gray line (red in the electronic
      version) the full geometric model.}\label{fig:mod}
  \end{center}
\end{figure}

\begin{figure}
  \begin{center}
    \plotone{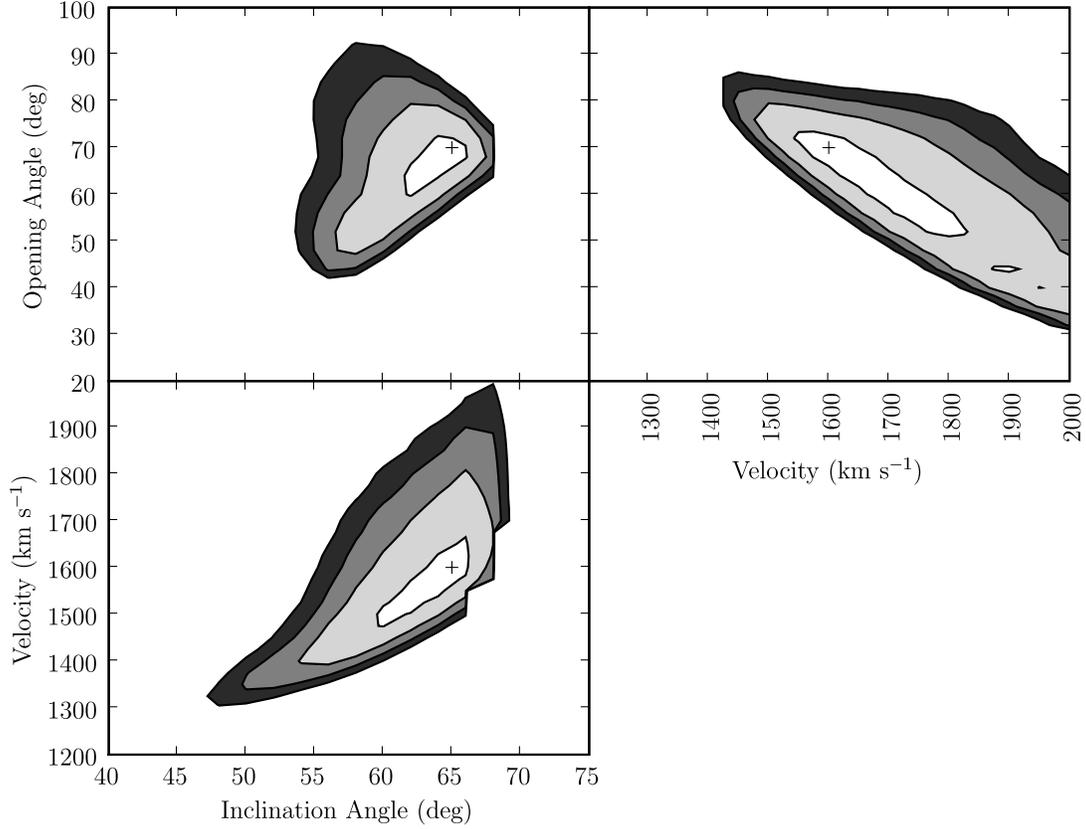}
    \caption{Geometric model uncertainty contours. The top left plot displays
      the uncertainty contours in opening and inclination angles with a fixed
      velocity of 1600 km~s$^{-1}$. The top right plot give the uncertainty
      contours in opening angle and velocity about a fixed inclination of 65
      deg. The bottom plot shows the uncertainty contours in velocity and
      inclination angle about a fixed opening angle of 70 deg. Contours in each
      plot give the 2, 3, 4, and 5 sigma deviations from the Gemini data. The
      plotted cross shows the position of the best fit value.}\label{fig:geomsig}
  \end{center}
\end{figure}

The resulting best fit model is displayed along with the Gemini data in Figure
\ref{fig:mod}. This model has an outflow velocity of 1600$^{+190}_{-90}$ \kms ,
an opening angle of 70 $\pm$ 8 degrees, and an inclination angle of
65$^{+3}_{-6}$ degrees relative to the line of sight.  The uncertainty values
on these parameters indicate the value at which a given parameter produces a 3
$\sigma$ deviation of the model velocity profile from the Gemini data, with the
other two parameters fixed to the best fit values. It is important to note that
the model parameters are not completely independent, and so these uncertainties
do not demarcate a region of ``acceptable'' fits in parameter space.  For
reference we include a set of cross-sectional uncertainty contours about the
best fit model parameters in figure \ref{fig:geomsig}. The Gaussian component
of the model has a FWHM of $\sim310$ km~s$^{-1}$. In this model 81 percent of
the flux is contributed by the bipolar conical outflow, and the remaining 19
percent by the lower velocity width Gaussian structure.  The flux ratio of the
receding outflow to the approaching outflow is 1.4. This simple model provides
an excellent fit to the [O~III]$\lambda$5007 line profile out to $\sim$1300
km~s$^{-1}$ from line center. Above this velocity the data show excess emission
relative to the model both at the red and blue extremes.

The model assumes an optically thin emitting region, as appropriate for
[O~III], and a uniform emission density.  The uniform emission
density is adopted to simplify the calculation, but provided there is no strong
density dependence with the angular separation from the central axis, any
radial dependence of the emission density will not change the velocity
structure of the model line. The bulk flow velocity is taken to be constant
over the outflow. This assumption, while assuredly an oversimplification,
provides a reasonable description of the available data.  It is worth noting
that in this formulation of the geometric model, the Gaussian component does
not have an explicit physical analog.  A number of physical structures could
produce a Gaussian velocity profile and we will consider a few possibilities in
later sections.

Given the simplicity of the high velocity width bipolar conical outflow plus
lower velocity width Gaussian model, we do not place great emphasis on the
detailed parameters of the model fit. Broadly, the large velocity width
requires a significant outflow velocity. The flux level asymmetry of red and
blue high velocity structures requires a non-uniform outflow. A flat velocity
profile, as observed in the high velocity width component, may also be produced
by a uniform spherical outflow or approximated by a radially directed disk
wind. However, the flux asymmetry of approaching and receding outflows observed
favors the spatially discrete structures of the bipolar conical outflow
considered above. Due to this asymmetry, the best fit profile of a spherical
outflow deviates from the Gemini data at the 9 sigma level.  A spherical shell
will produce an identical profile to a filled sphere because, as noted above,
the velocity profile of any outflow whose velocity is independent of radius, is
also independent of the radial density profile.  The large opening angle
indicates that the outflow that fits best in this model is not from a narrowly
collimated jet, but rather a broader outflow.  Any model which includes the
detailed hydrodynamics of the outflowing gas may deviate from the parameters
generated by the purely geometric fit considered here.  As such we present the
model parameters as descriptive and useful for consideration of the general
properties of the emitting region.

\section{DISCUSSION} \label{sec:dc}
\subsection {Variability}

There are two major modes by which the observed lack of strong variability in
the [O~III] emission might be produced in an accreting black hole system; a
steady accretion mode, or an emitting system in which the crossing time for
ionizing photons is large in comparison with the variability timescale of the
ionizing source. In the case of a variable ionizing source the emission
variability may also contain information on the density distribution of the
emitting gas.  The x-ray data described in \citet{Shih08} span nine years and
display a consistent count rate between observations. The four observations
described in the present work also display consistent [O~III]$\lambda$5007
equivalent widths between observations. For reference the light crossing path
bounded by the optical observations presented here is 0.4 pc. Both observations
are consistent with a long lived steady accretion rate. The sparse temporal
sampling of the data, however, does not allow tight constraints on shorter term
fluctuations in either the X-ray or [O~III] emission. Future x-ray observations
of the RZ2109 x-ray source, XMMU 1229397+07533, with finer temporal resolution
could be combined with the [O~III] variability limits presented here to
construct limits on the spatial scale of the emitting region.  Likewise
variability in the [O~III] emission under constant x-ray flux could yield
information on the outflow rate and density distribution.

The absence of strong variability does allow for the elimination of one class
of models as the source of the observed [O~III] emission.  Novae shells in the
local Milky War are a known source of [O~III]$\lambda$5007 emission
\citep{Downes2001}. The very brightest shell nova reach L([O~III]$\lambda$5007)
$\sim 10^{37}$ ergs and have velocity widths on the order of those observed in
RZ2109 \citep[][for example]{Downes2001,Iijima2011}.  In contrast to the
observations presented here, however, the [O~III] luminosity of these sources
decay by several orders of magnitude and velocity profile vary on timescales of
days to weeks. 

\subsection{Structure of Emission Region}

It is somewhat remarkable that the geometric model presented in Section
\ref{sec:gm} fits the [O~III] line profile as well at it does with simplistic
assumptions. There are, of course, a number of questions either not addressed
or raised by the model, including the spatial scale of the emitting system, the
observed flux asymmetry of the red and blue high velocity width component, the
failure of the model at the extreme high velocity wings, and the nature of the
low velocity Gaussian component.

The geometric model as presented is free of any spatial scaling.  We must look
to other means in order to place limits on the size of the outflow region.  A
firm lower limit may be set using the total emission from the high velocity
width outflow component of the [O~III]$\lambda 5007$ emission, adopting a pure
[O~III] gas at critical density for [O~III]$\lambda 5007$ at a temperature of
$10^{4}$ K \citep{Osterbrock89}, and employing the geometry parameters
found. With this minimal constraint the observed emission could originate from
a region with a minimum radius of a few $\times 10^{-3}$~pc.  The data do not
provide a clear upper limit on the size of the emitting region, and there is no
reason given the available data that the size of the [O~III]$\lambda 50007$
region could not be as large as the $\sim 10$ pc extent of the globular cluster
itself.

Asymmetries in the bipolar outflows are a well known feature of accreting black
hole systems. When a flux asymmetry is attributed to a differential obscuration
the usual pattern observed is a decrement in the flux of the receding flow
produced by the additional obscuring material along the greater line of sight
distance \citep{vanG87,Peterson00}. However, we observe the opposite effect in
the RZ2109, with the receding [O~III] emission line 40 percent more luminous
than the approaching flow.  If differential obscuration contributes to the
observed flux asymmetry, the obscuring media must be local to the cluster
system, such as dust-bearing gas at large radius resulting from an earlier
outflow of the system or interaction with a asymmetric ISM. The mechanism by
which such a local obscuring media may be produced is unclear.

A second known source of flux asymmetry in outflows is Doppler boosting
\citep{MR99,Fender2006}. In Doppler boosting, however, the flux of the
approaching outflow is amplified relative to the receding outflow, again
opposite of the asymmetry observed in RZ2109.  The different line of sight
travel times of outflows toward and away from the observer can also produce
asymmetric fluxes if coupled with a variable ionizing source.  However, the
flux asymmetry is the same in both our Keck and Gemini observations separated
by more than a year, which would be very difficult to understand in such a
scenario.  It therefore seems most likely that the somewhat greater flux on the
red side of the [O~III]$\lambda 4959$,$\lambda 5007$ line profile relative to
the blue side is a result of an intrinsic asymmetry in the source's emission
density. Such an asymmetry may be attributable to either a difference in the
mass of each outflow or the ionizing flux incident on them.

An expanding gas distribution such as nova shell or other remnant produced by
late stage stellar evolution and ionized by an external source may produce a
flux asymmetry across the velocity profile provided the gas is optically
thick. Such an arrangement would allow for a flux excess of the receding side
of the velocity profile relative to the approaching if the media was positioned
between the observer and the ionizing source.  For example, a late stage nova
shell modeled with a uniform density shell illuminated by an external x-ray
source produces an asymmetric velocity profile with a continuous blue to red
flux gradient whose slope is set by the position of the external x-ray source.
However, this type of flux asymmetry fails to provide an adequate description
of flat topped step function profile observed in the [O~III]$\lambda
4959$,$\lambda 5007$ line profiles high velocity width component.

The failure of the model at the most extreme velocities, seen in velocities in
excess of $\pm$1400 km~s$^{-1}$ in Figure \ref{fig:mod} is most likely
attributable to the simplifying assumption of constant bulk velocity in the
geometric model. If this is the case and the emission in these extreme velocity
wings is produced by the outflow closest to the source, prior to any
deceleration, with more sophisticated modeling techniques the extreme velocity
wings may provide another constraint on the spatial scale of the system.

While a model with a strong bipolar outflow can naturally account for broad
[O~III]$\lambda 4959$,$\lambda 5007$ emission, the origin of the narrower
Gaussian component is less clear. The observed FWHM of this lower velocity
width Gaussian peak is three times the FWHM of the spectral resolution of the
Gemini spectrum. Thus, the observed strongly peaked [O~III]$\lambda 5007$ line
profile is representative of its intrinsic shape. One implication of the
strongly peaked line profile is that rotation is ruled out for the origin of
the observed width of the line. It is worth noting that forbidden lines such as
[O~III] are optically thin, and the observed profile reflects the intrinsic
one, unlike the much more complex case of optically thick permitted lines such
as Balmer emission lines in AGN.  Therefore, for optically thin [O~III],
rotation naturally produces a double-peaked or flat-topped velocity profile
\citep[e.g.][]{Clark79}, and fails badly to match the centrally peaked profile
seen in the [O~III] emission. For example, the specific case of a rotating
Keplerian disk with an inverse square law emission density profile has a
velocity profile which deviates from the observed lower velocity width [O~III]
component at the $7\sigma$ level in a one spectral resolution element bin about
line center.  In fact, instead of showing a double-horned profile or very
flat-topped profile typical of rotation, the observed profile is more centrally
peaked than a Gaussian at about the $3\sigma$ level.  Therefore, rotation is
ruled out as the source of the velocity width of the lower velocity width
Gaussian component, contrary to the assumption of some modeling
\citep{Porter2010}.

The luminosity of the L([O~III])$\lambda 5007$ at a given velocity can also be
used to determine whether the velocity width of the line can be due to
gravitational motions or not.  Specifically, because [O~III]$\lambda 5007$ is a
forbidden line with a known critical density \citep{Osterbrock89}, we can
determine the maximum possible L([O~III])$\lambda 5007$ within a given volume
by adopting a pure [O~III] gas of uniform density at its critical density. We
can also compute the largest volume available sufficiently close to the black
hole to account for the observed velocity width due to gravitational motions
about the black hole.  \citet{Zepf08} carried out this calculation for the
broad velocity component of the [O~III]$\lambda 5007$, and showed it is many
orders of magnitude stronger than can be accounted for with gravitational
motions, and to thus require strongly outflowing material.

Here we apply this same calculation to the lower velocity width Gaussian
component which is seen much more clearly in our new Gemini data with its
greater spectral resolution and higher signal-to-noise. Adopting a pure [O~III]
gas of uniform density at its critical density \citep{Osterbrock89} and a
canonical black hole mass of 10\Msun, we find the maximum luminosity in a one
resolution element bin -170 km s$^{-1}$ from line center is of order 10$^{29}$
ergs s$^{-1}$. In contrast, the observed luminosity at this velocity is
$4.2\times 10^{35}$ ergs s$^{-1}$, and thus gravitational motions also fail to
account for the velocity width of the lower velocity width Gaussian peak.

It is worth noting that the above arguments do not strictly rule out rotation
of the emission line region, they do, however, require that the observed line
widths be produced by some mechanism other than rotation. A scenario where the
emitting region rotates in the plane of the sky, or with specific arrangements
of absorbing systems may be invoked, but in these cases the line width and
profile would be independent of any rotation.  Because the emitting region is
taken to be optically thin to the forbidden line emission observed, self
absorption or intermediate line absorption systems are unlikely. Gray body or
broad band absorption/scattering via an obscuring media is possible, however
the specific distribution of the material necessary to transform a rotational
line profile to that observed (a preferential and symmetric obscuration of the
emitting gas moving parallel/anti-parallel with the line of sight) is difficult
at best.

Although the specific calculation above is for a rotating sphere for
simplicity, adopting a dynamically hot model for the emitting material will
change the gravitational motions prediction far less than the five or more
orders of magnitude difference between it and the data.  The calculation also
adopts a stellar mass for the black hole in the globular cluster, based on
results in Z08. In that work, Z08 used the same type of volume calculation as
above for the broad component of the [O~III]$\lambda 5007$ emission, and showed
that it is not due to gravitational motions unless the black hole is nearly as
massive as the $\sim 10^6 \Msun$ cluster itself \citep[for a more detailed
  treatment see also][]{Porter2010}. The broad component then must be due to a
strong outflow.  Z08 suggest that observational evidence and theoretical
calculations \citep[and references therein]{Proga2007} indicate that strong
outflows are associated with high (L/L$_{\rm Edd}$) systems, which given the
observed L$_X$, implies a stellar mass for the black hole in RZ2109.

An alternative idea, that the source could be a tidal disruption or detonation
of a white dwarf by an intermediate-mass black hole, fails on several grounds
\citep[e.g.][]{Irwin2010}.  A disruption is unable to account for either the
high L([O~III]$\lambda 5007$) over an extended period of time or the long
duration of its high X-ray flux from at least the early 1990s ROSAT era through
2004 \citep{Maccarone07,Shih08}.  A detonation may account for the high velocity
outflow, but does not provide an explanation for why only oxygen lines are seen
in the spectrum when the detonation is powered by the fusion of oxygen into
iron. Furthermore, neither a disruption or detonation is able to account for
the more than a decade of consistently high L$_X$ followed by the recently
observed steep decline of L$_X$ by more than an order of magnitude
\citep{Maccarone10}.

With the current observations we have limited ability to address the energetics
of the outflow. Since we have no direct measurement of the temperature, we can
estimate the mass of [O~III] in the emitting region finding using a temperature
typical of [O~III] emission line. With an assumption of $\rm{T}=10^4$ K we find
an [O~III] mass of $7.3 \times 10^{-5}$ $\rm{M_{\sun}}$.  For temperatures from
$5 \times 10^3$--$10^6$ K the mass ranges from $1 \times 10^{-3}$ to $4 \times
10^{-6}$ $\rm{M_{\sun}}$.  Unfortunately since we have little to constrain
either the composition of gas, beyond that it is likely hydrogen depleted, or
the ionization fraction as no [O~II] lines are observed above the detection
threshold, we are unable to provide a concrete estimate of the total gas mass
of the emitting region. For a pure oxygen gas the total mass will be within a
factor of a few greater than the [O~III] mass, while a solar composition would
imply a total mass roughly two orders of magnitude greater. Using the 1600 \kms
velocity of the higher velocity component found in Section \ref{sec:gm} the
kinetic energy of the [O~III] is then 10$^{45}$ erg. This energy is 8 orders of
magnitude smaller than the available total accretion energy, assuming an
accretion efficiency of 10 percent and a 1 M$_{\sun}$ donor. If a solar
composition is assumed for the outflowing gas and all oxygen being doubly
ionized the systems kinetic energy is 6 orders of magnitude smaller than the
total accretion energy. It should be noted that the kinetic energy given here
is for the total observed outflow; it is unknown what fraction of the total
accretion event which the current outflow represents.

\section{SUMMARY}\label{sec:sum}
In this work we have presented a study of the equivalent width and velocity
profile of the [O~III]$\lambda \lambda 4959,5007$ emission line complex
associated with the black hole X-ray source in the extragalactic globular
cluster RZ2109. We have found little variation in the [O~III]$\lambda$5007
emission in our multiple observations over a time span of 467 days. The first
and last observations have a formal difference of $9 \pm 2\%$ with no evidence
for different velocities, and our upper limit in the variation among all our
observations is $27\%$. The [O~III]$\lambda 5007$ emission line is comprised of
two components; a lower velocity width component which is well characterized by
a Gaussian with FWHM 310 km~s$^{-1}$, and high velocity width component with a
width of 3200 km~s$^{-1}$ which we describe using a simple bipolar conical
outflow model.  The velocity width of the Gaussian component cannot be due to
rotation, and both components appear to require non-gravitational motions in
order to provide sufficient volume to produce the observed high luminosities in
the [O~III]$\lambda$5007 line.

\acknowledgements

MMS and SEZ wish to acknowledge support from NSF grants AST-0406891 and
AST-0807557. AK thanks NASA for support under Chandra grant GO0-11111A. KLR
acknowledges support from NSF Faculty Early Career Development grant
AST-0847109.  We thank the anonymous referee for the questions and critique
which helped to strengthen the paper. We also thank Jack Baldwin for helpful discussion.

\begin{deluxetable}{lcr}
  \tablewidth{0pt}
  \tablecaption{[O~III]$\lambda$5007 Equivalent Widths \label{tab:ew}}
  \tablehead{
    \colhead{Observation} & \colhead{eW (\AA)} & \colhead{Date (days)}
    }
  \startdata
  Keck & $30.9 \pm 0.7$ & 0.0 \\
  WHT & $42.6 \pm 2.5$ & 20.0 \\
  SOAR & $26.8 \pm 8.0$ & 434.1 \\
  GMOS & $34.0 \pm 0.4$ & 467.0 \\ \hline
  \enddata
\end{deluxetable}

\end{document}